\documentclass[a4paper,11pt]{article}
\pdfoutput=1 

\usepackage{jinstpub} 
\usepackage{parskip}
\usepackage{float}
\usepackage{bold-extra}
\usepackage{amsmath}
\usepackage{amssymb}
\usepackage{subfigure}
\usepackage{times, txfonts}
\usepackage{mathtools}
\usepackage{lineno}
\usepackage{hepunits} 
\usepackage{amsmath}
\title{\boldmath Experiences from the RPC data taking during the CMS RUN-$2$}

\author[z,1]{M A Shah,\note{Corresponding author.}}

\author[a]{A. Samalan}
\author[a]{,M. Tytgat}
\author[a]{,N. Zaganidis}
\author[b]{,G.A. Alves}
\author[b]{,F. Marujo}
\author[c]{,F. Torres Da Silva De Araujo}
\author[c]{,E.M. Da Costa}
\author[c]{,D. De Jesus Damiao}
\author[c]{,H. Nogima}
% ...

\affiliation[a]{Ghent University, Dept. of Physics and Astronomy, Proeftuinstraat 86, B-9000 Ghent, Belgium}
\affiliation[b]{Centro Brasileiro Pesquisas Fisicas, R. Dr. Xavier Sigaud, 150 - Urca, Rio de Janeiro - RJ, 22290-180, Brazil}
\affiliation[c]{Dep. de Fisica Nuclear e Altas Energias, Instituto de Fisica, Universidade do Estado do Rio de Janeiro, Rua Sao Francisco Xavier, 524, BR - Rio de Janeiro 20559-900, RJ, Brazil}
% ...
\affiliation[z]{National Centre for Physics, Quaid-i-Azam University, Islamabad, Pakistan.}
% ...

\emailAdd{mashah@cern.ch}

\abstract{The CMS experiment recorded 177.75 \invfb of proton-proton collision data during the RUN-$1$ and RUN-$2$ data taking period. Successful data taking at increasing instantaneous luminosities with the evolving detector configuration was a big achievement of the collaboration. The CMS RPC system provided redundant information for the robust muon triggering, reconstruction, and identification. To ensure stable data taking, the CMS RPC collaboration has performed detector operation, calibration, and performance studies. Various software and related tools are developed and maintained accordingly. In this paper, the overall performance of the CMS RPC system and experiences of the data taking during the RUN-$2$ period are summarised.}

\keywords{Resistive-plate chambers}

\collaboration[c]{on behalf of the CMS collaboration}
\proceeding{15$^{\text{th}}$ Workshop on Resistive Plate Chambers and related detectors (RPC2020) 
\\
  Roma, Italy\\
  10-14 February 2020}
\usepackage{float}%
\usepackage{graphicx, subfigure}
\begin{document}
\maketitle

\flushbottom
\section{Introduction}
\label{sec:intro}

One of the key features of the CMS (Compact Muon Solenoid) experiment \cite{b} is its extensive muon system \cite{d}. As a powerful handle to the signature of interesting events, the triggering and reconstruction capabilities for muons are very important. The CMS muon system exploits three different gaseous technologies, namely, Drift Tubes (DT) in the barrel (central) region, Cathode Strip Chambers (CSC) in the endcap (forward) region, and Resistive Plate Chambers (RPC) \cite{c} in both the barrel and endcap, covering up a pseudo-rapidity region of $|\eta| < 2.4$, where RPCs are installed up to $|\eta| < 1.9$. The muon system has the key functions of muon triggering, transverse momentum measurement, muon identification, and charge determination.


\section{CMS RPC Operation and Performance During RUN-$2$}

During RUN-$1$ and RUN-$2$, the CMS detector has recorded proton-proton collisions data amounting to 177.75 \invfb with 150.26 \invfb data at $\sqrt(s)~$=$~13$ \TeV~during RUN-$2$ only. During the whole period, the RPC system has contributed very efficiently in the data collection. The total accumulated charge for CMS RPC was measured to be 2.3 mC/cm$^2$ for barrel and $7.5$ mC/cm$^2$ for the endcap. The fraction of luminosity loss due to RPC problems during the entire RUN-$2$ was just $0.15\%$.

The CMS RPCs are used mainly as triggering detectors. The RPC system consists of 1056 double gas gap chambers made of high pressure laminate plates (HPL, commonly known as Bakelite) with a bulk resistivity in the range of $10^{10} - 10^{11}\, \Omega\cdot$cm. The performance of RPCs depends on the usage of a proper working gas mixture. To operate in avalanche mode the CMS RPCs are using a composition of $3$ gases: 95.2\% freon ($C_2H_2F_4$) to enhance an ionization caused by the incident particle, 4.5\%  isobutane ($iC_4H_{10}$) used as a quencher gas to reduce streamer formation, and 0.3\% sulphur hexafluoride ($SF_6$) to control the background electrons produced from secondary ionization and clean the signal.

\subsection{RPC Efficiency and Cluster Size Stability}

Important parameters for the RPC system performance monitoring are the RPC hit registration efficiency and cluster size (CLS), where CLS refers to the number of adjacent strips fired in response to the passage of charged particles. The RPC hits coordinates are calculated in the geometrical center of the formed clusters of fired strips. Keeping the optimal CLS can improve proper estimation of the bending angle of the muon trajectory. To follow the muon triggering requirements the cluster size of the RPC hit should be kept not more than $3$ strips. The proper calibration of the detector is based on the analysis of efficiency and cluster size dependences on the applied high voltage.  The HV scan is taken at effective, equidistant voltages in the working range of $[8600, 9800]$ V. The collected data are being analyzed to evaluate the optimal high voltage working points (HV\_WP). More details about the RPC HV scan methodology can be found in \cite{e, f}. The results of HV scans up to 2017 with comparison to previous years can be found in \cite{rogelio}.

\begin{figure}[!htb]
\centering
\subfigure[]{%
\includegraphics[width=.41\textwidth]{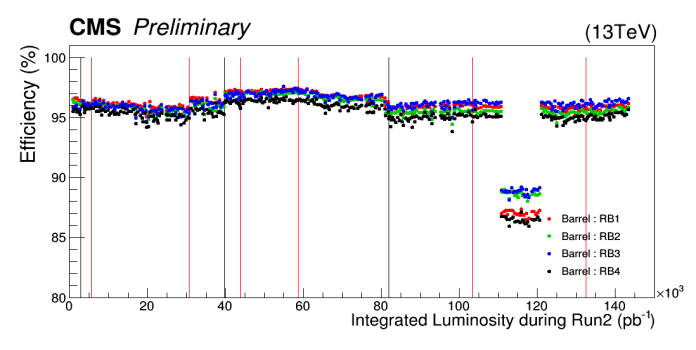}
\label{fig:subfigure1}}
\quad
\subfigure[]{%
\includegraphics[width=.41\textwidth]{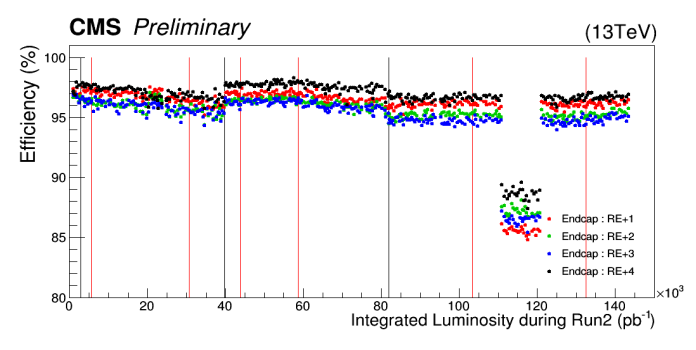}
\label{fig:subfigure2}}
\caption{RPC efficiency vs integrated luminosity during RUN-$2$ for barrel in (a) and endcap in (b).}
\label{fig:1a}
\end{figure}

\begin{figure}[!htb]
\centering
\subfigure[]{%
\includegraphics[width=.42\textwidth]{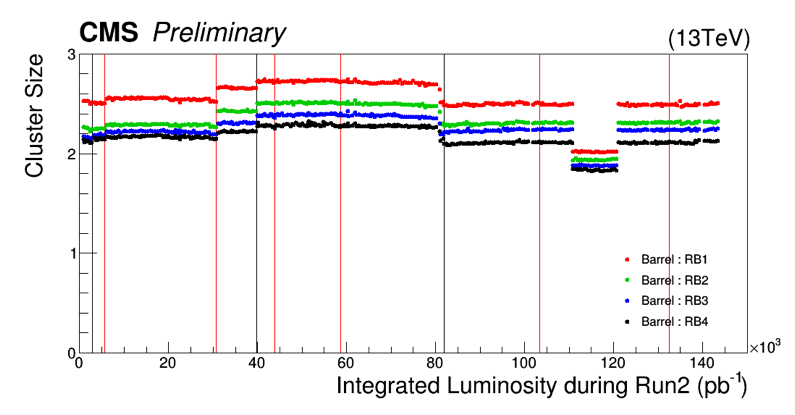}
\label{fig:subfigure1}}
\quad
\subfigure[]{%
\includegraphics[width=.42\textwidth]{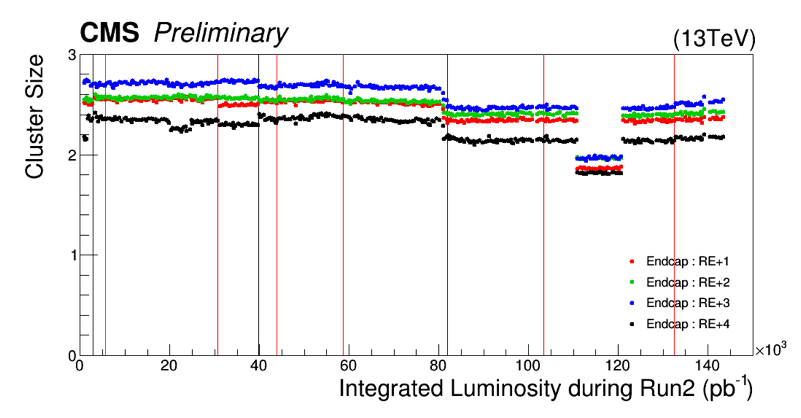}
\label{fig:subfigure2}}
\caption{RPC cluster size vs integrated luminosity during RUN-$2$ for barrel in (a) and endcap in (b).}
\label{fig:1b}
\end{figure}

Figure \ref{fig:1a} represents RUN-$2$ efficiency history and Fig. \ref{fig:1b} shows cluster size history for barrel and endcap respectively. Each point corresponds to an average efficiency or cluster size per station in a given LHC fill. Data points with low statistics or temporary problems are excluded from the distributions. The x-axis shows the integrated luminosity and the y-axis shows average efficiency or cluster size for the detector part under study. Red lines are the planned technical stops (TS) and the grey ones - Year-End-Technical stops (YETS). The trend of the curves follows the changes in the applied high voltage working points and changes in the isobutane concentration in the used gas mixture. The drop in the efficiency and CLS during 01. Aug. 2018 - 19. Aug. 2018 is caused by a known configuration setting problem \cite{rpc2018}.

\begin{figure}[!htb]
\centering
\subfigure[]{%
\includegraphics[width=.41\textwidth]{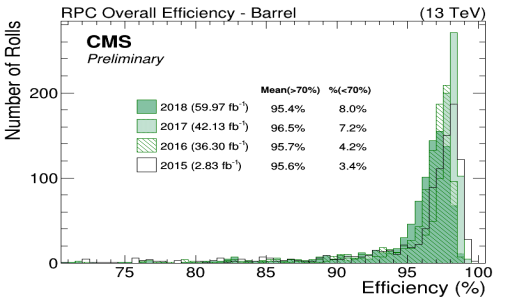}
\label{fig:subfigure1}}
\quad
\subfigure[]{%
\includegraphics[width=.41\textwidth]{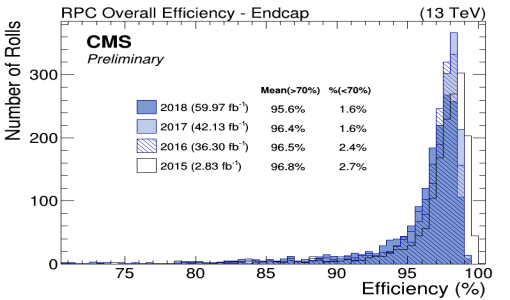}
\label{fig:subfigure2}}
\caption{RPC efficiency distributions during RUN-$2$ for barrel in (a) and endcap in (b).}
\label{fig:1c}
\end{figure}

In 2016, because of higher Isobutane concentration (5.3\%), efficiency was lower as the HV\_WP were not changed to compensate for the wrong gas mixture. After the deployment of the modified WP in September 2016, the efficiency increased slightly by ~1\% and cluster size increased sharply. Gas concentration was back at 4.5 \% in 2017 but the WP were not changed. The efficiency remained unchanged because the detectors were running in the plateau of the sigmoid curve, however an increase in the cluster size has been observed. New WP had been deployed by the end of 2017, which led to a slight decrease in the efficiency but a sensible reduction in the cluster size. Efficiency distribution of RPC for barrel and endcap during RUN-$2$ is shown in Fig. \ref{fig:1c}. The overall efficiency during the RUN-$2$ were kept  around 96\%.

\subsection{RPC Currents}
The ohmic current of RPC is defined as current with no beam, up to around 7000 V, in the range where there is no contribution of gas amplification and the current follows the ohmic law. The ohmic current values are monitored at 6500 V. Cosmic current is defined as current without beam, at working point voltage, in the region of the gas amplification. 

\begin{figure}[!htb]
\centering
\includegraphics[width=.82\textwidth]{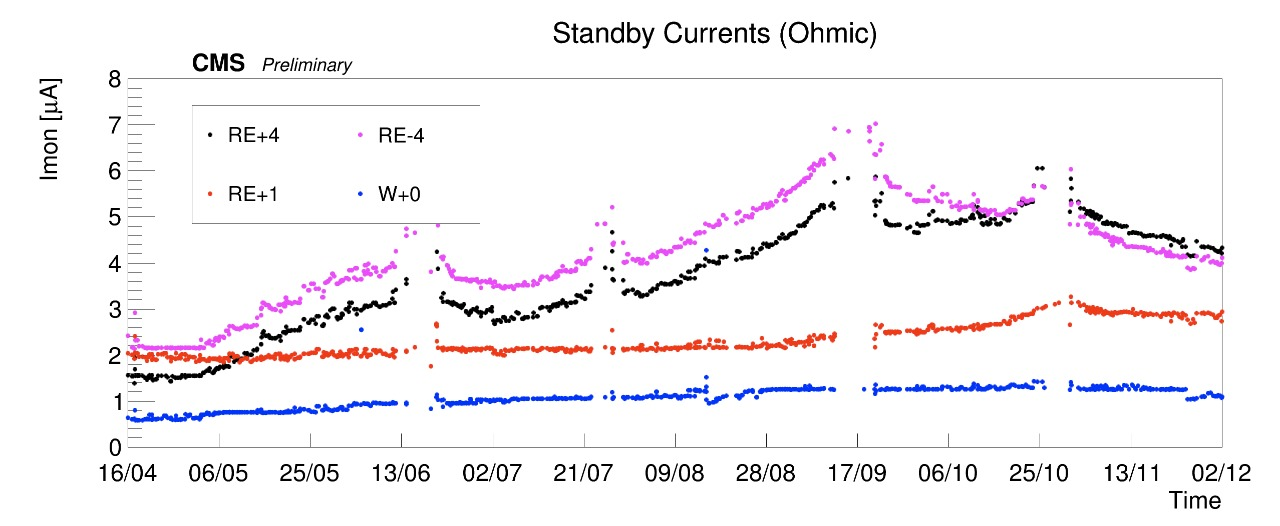}
\caption{Ohmic current history of W+0, RE+1, RE+4 and RE-4.}
\label{fig:3a}
\end{figure}

Figure \ref{fig:3a} shows currents measured in four RPC stations, W+0 in the barrel and RE+1, RE+4, RE-4 in the endcap. The measured currents are shown as a function of time. From the beginning of 2018 to September, a higher ohmic current has been observed in RE-4 compare to RE+4. After doubling the gas flux in RE-4 by middle of September, a faster reduction of currents has been observed with respect to RE+4. The increase of ohmic currents has been observed directly correlated with the background. In the low background regions such as W+0, a very slow increase in the ohmic current has been observed. In RE+1 and W+0 the background rate is less than 10 Hz/cm$^2$ and both have similar gas flows (0.7 volume exchange per hour (v/h) and 0.6 v/h respectively), while in RE4 the background rate is about 40 Hz/cm$^2$ and the gas flow is 1.1 v/h.

\begin{figure}[!htb]
\centering
\includegraphics[width=.75\textwidth]{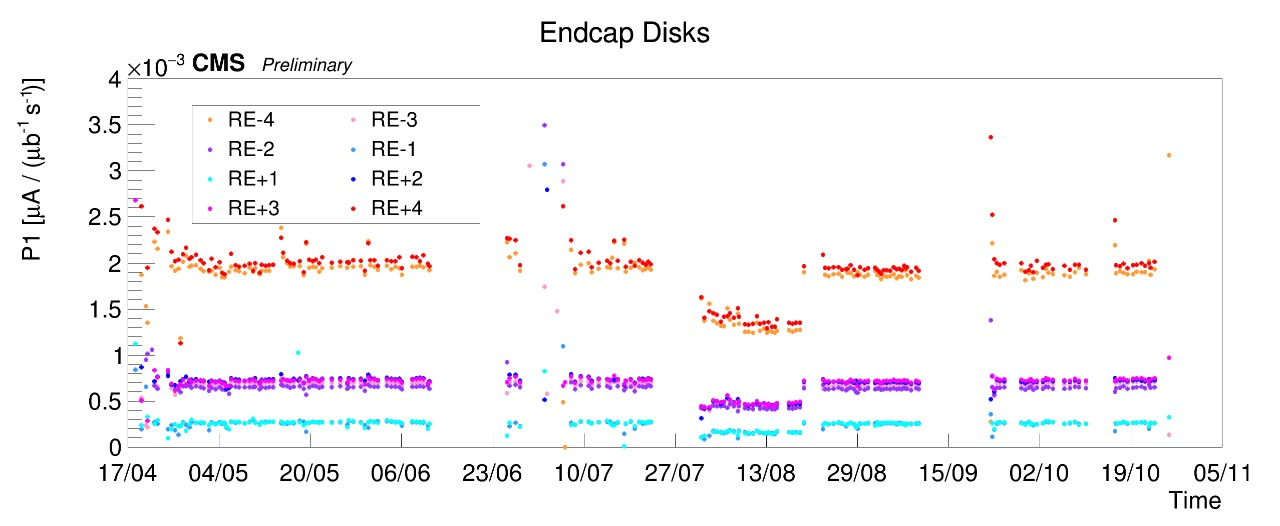}
\caption{RPC cosmic current history of W+0, RE+1, RE+4 and RE-4.}
\label{3bc}
\end{figure}

The RPC currents depend linearly on the instantaneous luminosity \cite{r5}. For each LHC fill the linear distributions were fit to a linear function in order to obtain the slope ($P_{1}$) also known as physics current $(i ~$=$~ P_{1} \times L)$. Due to the nature of the linear fit, $P_{0}$ (offset) absorbs the cosmic current (offset + ohmic + gas gain). The slopes ($P_{1}$) as a function of time for the endcap stations are shown in Fig. \ref{3bc}. The slope of the RPC currents distribution is stable in time. The changes in the middle of August are due to different applied HV working points. Endcap stations, located at equal distances from the interaction point along the beam pipe, have similar slopes ($P_{1}$ values). They also have similar rates \cite{r5}. No increase due to integrated luminosity is observed for the slopes for the entire year.

\begin{figure}[!htb]
\centering
\includegraphics[width=.75\textwidth]{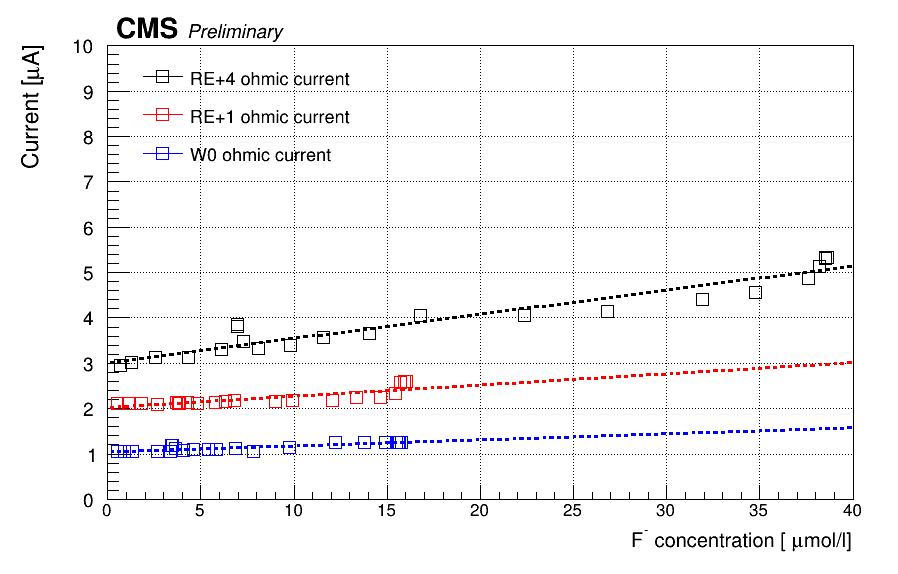}
\caption{Ohmic current as a function of HF concentration.}
\label{fig:3c}
\end{figure}

Hydrogen Fluoride (HF) is produced in the gas under high electrical discharge, which has a high chemical reactivity and electrical conductivity. Therefore, it is supposed that HF can be a source of inner detector surface damaging and relative ohmic current increase which accelerate detectors ageing. The HF measurements have been performed using an ion-selective electrode (ISE), which is a transducer (or sensor) that converts the activity of a specific ion $[F^{-}]$ dissolved in a solution into an electrical potential. The measurements have been performed at the gas exhaust of 3 regions: W+0 in the barrel and RE+1, RE+4 in the endcap. The ohmic currents as a function of HF concentration is shown in Fig. \ref{fig:3c}. RE+1 and W+0, have a similar HF concentration, gas flow (0.7 v/h and 0.6 v/h ) and background (less than 10 Hz/cm$^2$). In RE+4 the amount of HF accumulated is around 2 times higher with higher background (40 Hz/cm$^2$) and the gas flow is 1.1 v/h, 2 times more than W+0 and RE+1. There is a clear linear dependence between the ohmic current and HF concentration which implies that HF trapped in the gap may form a thin conductive layer. HF can be efficiently removed by fine tuning the gas flow depending on the background rate.

\section{Conclusion}
CMS RPCs have been operating very successfully during RUN-$2$.  After 9 years of LHC running with increasing instantaneous luminosity and several years from the end of RPC construction, the detector performance is within CMS specifications and stable without any significant degradation. A reversible ohmic current increase was observed in the most exposed regions. Fine tuning of the gas flux is mandatory for further detector operation. No significant issues were found for running up to high luminosity scenarios of LHC RUN-3.

\acknowledgments We would like to thank especially all our colleagues from the CMS RPC group and L1 muon trigger group for their dedicated work to keep the stable performance of the RPC system. We wish to congratulate our colleagues in the CERN accelerator departments for the excellent performance of the LHC machine. We thank the technical and administrative staff at CERN and all CMS institutes.



\begin{thebibliography}{99}
\bibitem{b}
CMS Collaboration, \emph{The CMS experiment at the CERN LHC, }J. Instrum. 3 (2008) S08004.
\bibitem{d}
CMS Collaboration, \emph{The CMS muon project$\colon$ Technical Design Report.} 1997. CERN-LHCC-97-032, CMS-TDR-003.

\bibitem{c}
R. Santonico and R. Cardarelli, \emph{Development of resistive plate counters,} Nucl. Instrum. Meth. 187 (1981) 377.



\bibitem{e} 
C. Camilo [CMS Collaboration], \emph{The CMS Resistive Plate Chambers system-detector performance during 2011,} WSPC (2012), 10.1142/9789814405072-0068.
\bibitem{f}
M. Abbrescia [CMS Collaboration], \emph{Cosmic ray test of double-gap resistive plate chambers for the CMS experiment}, Nucl. Instrum. Meth. A550 (2005) 116.
\bibitem{rogelio}
Rogelio Reyes Almanza [CMS Collaboration], \emph{High voltage calibration method for the CMS RPC detector}, JINST,14,C09046.
CMS Collaboration, \emph{Performance of CMS muon reconstruction in pp collision events at $\sqrt{s} ~$=$~ 7 TeV$,} 2012 JINST 7 P10002, [arXiv:1206.4071].
\bibitem{r5}
CMS Collaboration, \emph{RPC Detector Performance Results for 2016 and 2017} ; CMS-DP-2018-001.

\bibitem{rpc2018}
M A Shah and R Hadjiska [CMS Collaboration], \emph{The CMS RPC Detector Performance and Stability during LHC RUN-2} ;  JINST 14 (2019) no.11, C11012.

\end{thebibliography}
\end{document}